\documentclass[aps,preprintnumbers,prl,twocolumn,superscriptaddress]{revtex4} 
\usepackage{slashed}
\usepackage{graphicx}
\usepackage{color}
\usepackage{natbib}

\begin{document}

\preprint{ACFI-T15-01}

\title{Catalysis of Electroweak Baryogenesis via Fermionic Higgs Portal Dark Matter}

\author{Wei  Chao}
\email{chao@physics.umass.edu}
\author{Michael J. Ramsey-Musolf$^{2,}$}
\email{mjrm@physics.umass.edu}
\affiliation{ Amherst Center for Fundamental Interactions, Department of Physics, University of Massachusetts-Amherst
Amherst, MA 01003 USA\\
$^2$California Institute of Technology, Pasadena, CA 91125 USA}

\begin{abstract}

We investigate catalysis of electroweak baryogenesis by fermionic Higgs portal dark matter using a two Higgs doublet model augmented by vector-like fermions. The lightest neutral fermion mass eigenstate provides a viable dark matter candidate in the presence of a stabilizing symmetry $Z_2$ or gauged U(1$)_D$ symmetry. Allowing for a non-vanishing CP-violating phase in the lowest-dimension Higgs portal dark matter interactions allows generation of the observed dark matter relic density while evading direct detection bounds. The same phase provides a source for electroweak baryogenesis. We show that it is possible to obtain the observed abundances of visible and dark matter while satisfying present bounds from electric dipole moment (EDM) searches and direct detection experiments. Improving the present electron (neutron) EDM sensitivity by one (two) orders of magnitude would provide a conclusive test of this scenario.


\end{abstract}

\maketitle
\noindent {\bf Introduction} 
Precise cosmological observations have confirmed the existence of non-baryonic cold dark matter $\Omega_{DM} h^2 =0.1187\pm 0.0035$\cite{Ade:2013zuv,Komatsu:2010fb}. How dark matter interacts with the Standard Model (SM) particles remains unknown. The discovery of the Higgs boson opens a new avenue for both probing the dark universe and modeling its dynamics.  In particular, Higgs portal dark matter~\cite{Patt:2006fw,Kim:2006af,MarchRussell:2008yu,Kim:2008pp,Ahlers:2008qc,Feng:2008mu,Andreas:2008xy,Barger:2007im,Barger:2008jx,Kadastik:2009ca,Gonderinger:2009jp,Piazza:2010ye,Arina:2010an,Kanemura:2010sh,Englert:2011yb,Low:2011kp,Djouadi:2011aa,Kamenik:2012hn,Gonderinger:2012rd,Lebedev:2012zw,LopezHonorez:2012kv,Okada:2012cc,Djouadi:2012zc,Bai:2012nv,Englert:2013gz,Bian:2013wna,Chang:2013lfa,Khoze:2013uia,Okada:2013bna,Chao:2014ina,Cai:2014hka}, which extends the SM with a SM gauge singlet  that only interacts with the Higgs boson, provides one of the simplest realizations of this idea. 

It is possible that Higgs portal interactions may also provide an explanation for another unsolved problem in cosmology: the origin of the baryon asymmetry of the Universe (BAU).  Combining data from Planck~\cite{Ade:2013zuv},  WMAP~\cite{Komatsu:2010fb} and  large scale structure measurements, one has $Y_B \equiv \rho_B /s=(8.59\pm0.11)\times 10^{-11}$, where $\rho_B$ is the baryon number density and $s$ is the entropy density. Assuming that the Universe was matter-antimatter symmetric at the end of the inflationary epoch, it is likely that interactions involving beyond the Standard Model fields generated the BAU during the subsequent cosmological evolution. To generate the observed BAU, these early universe interactions must satisfy the three Sakharov criteria~\cite{Sakharov:1967dj}: (1) baryon number violation; (2) C and CP violation; (3) a departure from the thermal equilibrium (assuming CPT conservation).  

Fulfilling the criteria requires physics beyond the SM (BSM). Among the most attractive scenarios is electroweak baryogenesis~\cite{Morrissey:2012db,Cohen:1993nk,Trodden:1998ym,Riotto:1998bt,Riotto:1999yt,Quiros:1999jp,Dine:2003ax,Cline:2006ts}, as the relevant BSM interactions can be tested experimentally through direct searches for new particles at high energy colliders and probes of CP-violation (CPV) through searches for permanent electric dipole moments (EDMs)\cite{Engel:2013lsa,Pospelov:2005pr}. It is interesting to ask whether Higgs portal interactions responsible for the observed dark matter relic density may also generate the BAU through electroweak baryogenesis (EWBG). 

In this paper, we explore this question with a simple Higgs portal realization involving fermionic dark matter. Denoting the latter by $\chi$, the lowest dimension Higgs portal interactions are non-renormalizable and have the form~\cite{LopezHonorez:2012kv}
\begin{eqnarray}
{\alpha \over \Lambda} \bar \chi \chi (H^\dagger H) + {\beta \over \Lambda } \bar \chi i\gamma_5 \chi (H^\dagger H ) \; , \label{old}
\end{eqnarray}
where $\alpha,~\beta$  are effective couplings and $\Lambda$ is the scale associated with the effective theory. Both interactions may yield an annihilation cross section consistent with the observed relic density under a thermal dark matter scenario. However, the first term with coefficient $\alpha$ is severely constrained by dark matter direct detection experiments XENON100~\cite{Aprile:2012nq} and LUX~\cite{Akerib:2013tjd}, making it an unlikely candidate for dark sector physics. On the other hand, the sensitivity of direct detection  searches to the interaction proportional to $\beta$ is considerably weaker, due to the associated velocity suppression for a non-relativistic $\chi$. In what follows, we show that CPV Higgs portal interactions may lead to $|\beta| >> |\alpha| \sim 0$, as needed for consistency with the observed relic density and direct detection constraints,  while also providing a sufficient source for the BAU through EWBG.

\noindent{\bf Model}
We work in the Type-I two Higgs doublet model~\cite{Branco:2011iw}, wherein the second Higgs doublet has no Yukawa interactions with the SM fermions.  By introducing a pair vector-like fermions, $\psi_{L,R}$, transforming as $(1, ~2,~1/2)$ and  $\chi_{L,R}$ transforming as $(1,~1,~0)$, the new Yukawa interaction and mass term can be written as
\begin{eqnarray}
\sum_f\overline{f_L}^{} M_f^{}  f_R^{} + \sum_i \left( y_i^{} \overline{\psi_L^{} } H_i^{} \chi_R^{} + y_i^\prime \overline{\chi_L^{} } H_i^\dagger \psi_R^{} \right)  + {\rm h.c.} \label{boss}
\end{eqnarray}
where $f\equiv \psi,\chi $. Clearly,  $f_{L,R}$ may couple to the SM  leptons through Yukawa interactions, precluding the possibility of a stable neutral fermion. To ensure stability, one may impose a $Z_2$ symmetry, in which new fermions are odd, while all the other fermions are even.  Alternatively, one may also introduce a local $U(1)_D$ gauge symmetry, in which only $\psi_{L,R}$ and $\chi_{L,R}$ have non-vanishing but identical charges. 
The new fermion mass matrix can be written as
\begin{eqnarray}
\overline{\left( \matrix{\chi_L&\psi_L}\right) }\left( \matrix{ M_\chi& y_1^\prime v_1 + y_2^ \prime v_2 \cr y_1 v_1 + y_2 v_2 & M_\psi} \right) \left( \matrix{ \chi_R \cr \psi_R}\right) \; . \label{scp}
\end{eqnarray}
As observed in Ref.~\cite{Chao:2014dpa} the interactions (\ref{scp}) contain CP-violating phases that cannot be rotated way by field redefinition. 

The mass matrix in Eq.~(\ref{scp}) can be diagonalized by performing separate flavor rotations on the left- and right-handed components of the new neutral fermions, {\em viz}
\begin{equation}
\chi_L =  \mathcal{U}_{L\, 1i}\ \xi_{L\, i}\ ,\qquad
\psi_L  =  \mathcal{U}_{L\, 2i}\ \xi_{L\, i}
\end{equation} 
where $\xi_{1,2}$ are the two mass eigenstates and where similar rotations are performed on the right-handed fields with matrix $\mathcal{U}_R$. In the limit that these matrices are nearly diagonal, one has $\xi_1\approx\chi$ and $\xi_2\approx \psi$, where $\psi=\psi_L + \psi_R $ and $\chi =\chi_L + \chi_R$. Assuming $\xi_1\approx\chi$ is the lighter of the two states and does not decay on cosmologically relevant time scales (see below), 
 one may obtain the effective dark matter-Higgs interactions  by integrating out $\xi_2$ and the charged component of the vector-like fermion doublet:
\begin{eqnarray}
\sum_{i=1}^2 \left| y_i^\prime y_i^{}\over M_\psi \right | \left\{  \cos \delta_i  \bar \chi \chi + i \sin \delta_i  \bar \chi \gamma_5 \chi \right\} h_i^\dagger h_i^{} +\cdots  \; ,\label{effcoupling}
\end{eqnarray} 
where $\delta_i ={\rm Arg} (y_i y_i^\prime /M_\psi )$, $h_i^T=(H_i^+, H_i^0-v_i)$,  and the \lq\lq $+\cdots$" denote interactions containing $h_1^\dagger h_2 $ and (or) $h_2^\dagger h_1$ or terms suppressed by $\psi-\chi$ mixing.
Comparing with (\ref{old}) we observe that $\alpha\sim \cos\delta_i$ and $\beta\sim\sin\delta_i$ and $\Lambda=M_\psi^{}$. For sufficiently large $|\sin\delta_i|$, one may in principle satisfy the dark matter relic density and direct detection constraints while providing a sufficiently strong CPV source for EWBG.



\noindent {\bf Dark Matter} 
For illustrative purposes, we assume the second Higgs doublet is superheavy, which is consistent with the current LHC measurements, so that the effective interactions that dominate the annihilation of the DM are $(\bar \chi \{1,\gamma_5\} \chi ) h_1^\dagger h_1$.  A more comprehensive study of the dark matter phenomenology in the 2HDM will be given a forthcoming paper~\cite{2hdmpdm}. For the annihilation in the early universe responsible for setting the thermal relic density, dark matter is moderately relativistic. For direct detection, on the other hand, the dark matter particles are non-relativistic, with typical velocity  $v\sim 10^{-3}$ in the galactic halo~\cite{Bertone:2004pz}.  As a result, the parity-violating effective coupling contributes to dark matter annihilation but gives a velocity suppressed contribution to the dark matter-nucleus scattering cross section. It is, thus, possible to choose values of $\sin\delta_i/M_\psi^{}$ that allow for sufficiently efficient annihilation as needed to obtain the observed relic density while evading the stringent spin-independent direct detection bounds.

\begin{figure}[ht]
\begin{center}
\includegraphics[width=0.38\textwidth]{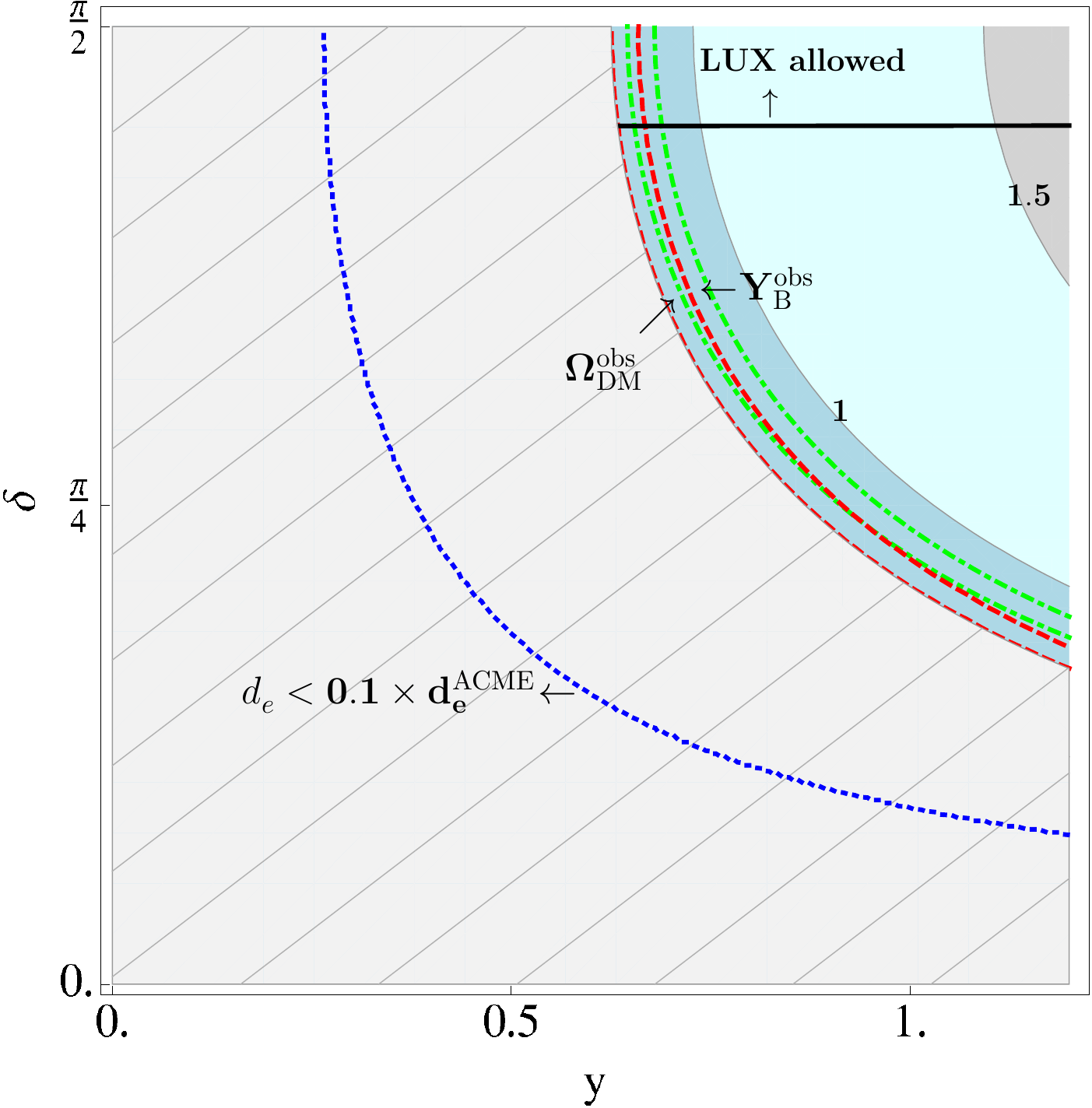} 
\end{center}
\caption{Contours of constant baryon asymmetry $Y_B$ in units of $10^{-10}$  in the $y-\delta $ plane (see text for parameter definitions). Dashed green band indicates a value of $Y_B$ consistent with the observed asymmetry, while the red dashed band indicates parameter region consistent the observed dark matter relic density. Grey hatched region leads to an over abundance of dark matter. Dark matter direct detection constraints from the LUX experiment~\cite{Akerib:2013tjd} allow the region above the solid black line. A future search for the electron EDM with an order of magnitude greater sensitivity than the ACME result~\cite{Baron:2013eja} would probe the parameter space above the blue dashed line.}
\label{BAU}
\end{figure}

We illustrate this feature in Fig. \ref{BAU}, choosing $y_1=ye^{i\delta }$ with $\delta$ being the only CP phase, $y_1^\prime =y_2 =0.5$ , $y_2^\prime=0$, $\tan \beta=15$, $M_\chi=300~{\rm GeV}$ and $M_\psi=1500~{\rm GeV}$. In this case the heavy-light ($\chi$-$\psi$) mixing is  ${\cal O}(10^{-2})$ and can be safely neglected. The resulting region between the red dashed lines gives a dark matter relic density consistent with Planck and WMAP measurements at $95\%$ C.L. . The grey hatched region leads to an over-abundance of DM, while the area to the upper right of the red-dashed curves corresponds to under-saturating the relic density. Regions below the black solid line are excluded by the LUX result~\cite{Akerib:2013tjd}.


\noindent {\bf Electric dipole moments}  The CP-odd Yukawa couplings given in (\ref{boss}) generate an elementary EDM {\it via} two-loop Barr-Zee diagrams~\cite{Barr:1990vd}. For illustrative purposes, we will work in the limit that the masses of the remaining neutral scalars and charged scalars are sufficiently heavy that the dominant contributions arise from the exchange of a $W^+W^-$ pair. CP-violation enters  the contribution through the relative phase of left- and right- handed charged currents.  A direct calculation gives the following result for the electron EDM 
\begin{eqnarray}
d_e\approx d_e^{(2l)} s_W^{-2}  {\rm Im } ({\cal U}_{L2i }^{} {\cal U}_{R2i}^*) {m_{\psi^+} m_i \over m_W^2 }  f_{WW} (r_1,r_2)
\end{eqnarray}
where $d_e^{(2l)} \approx 2.5 \times 10^{-27} e\cdot{\rm cm}$, $s_W =\sin \theta_W$ with $\theta_W$ being the weak mixing angle, 
$r_1= (m_{\psi^+}/m_W)^2$ and $r_2=(m_i /m_W)^2$,  and the loop function $f_{WW}(x, y)$ is given in Ref.~\cite{Giudice:2005rz}.  
An analogous expression applies for the quark EDMs that give rise to the EDM of the neutron. 

Present constraints obtained by the ACME collaboration~\cite{Baron:2013eja} yield the most stringent limit on the electron EDM: $d_e=(-2.1\pm 4.5)\times 10^{-29}$ $e$ cm . This bound is nearly two orders of magnitude more stringent than the current neutron EDM limit\cite{Baker:2006ts}. However, given the order-of-magnitude larger value for the light quark Yukawa couplings, the electron and neutron EDM constraints on this scenario differ by only by roughly a factor of ten. Nonetheless, the bounds from $d_e$ are presently too weak to impact the parameter space shown in Fig. \ref{BAU} largely because only the $W^+W^-$ graph contributes to the EDM (the situation is analogous to the two-loop chargino-neutralino EDM contributions in ``split  supersymmetry"   \cite{Giudice:2005rz,Li:2008kz,Li:2008ez}).  
Planned future improvements in EDM search sensitivities would change this situation. To that end, we show in Fig. \ref{BAU} the parameter space  that would be probed by a possible future $d_e$ search that is an order of magnitude more sensitive than the present ACME bound (see~\cite{Chao:2014dpa} for a detailed analysis).  Regions to the right of blue dotted line would be excluded  should a null result be obtained with this sensitivity. The inputs are the same as these in making the dark matter curves.

\noindent {\bf Baryogenesis} 
The three Sakharov conditions are realized in the following way. First,  a strongly first order electroweak phase transition, which provides a departure from equilibrium~\cite{Cline:1996mga,Fromme:2006cm,Dorsch:2013wja}, is induced by the two Higgs doublet potential. The phase transition occurs as bubbles of broken electroweak symmetry, which nucleate and expand in a background symmetric phase, fill the Universe.  Second, CP violation arises from the Yukawa couplings of the new fermions. A non-zero CP-asymmetric charge density is produced via the CP-violating interactions at  the walls of the expanding bubbles, where the Higgs vacuum expectation value (VEV) is space-time dependent. This charge density diffuses ahead  the expanding bubbles and is transformed into charge densities of other species via inelastic interactions in the plasma, resulting in a net density of left-handed  fermions $n_L$.  Third, the weak sphaleron processes, which are unsuppressed in the symmetric phase, violate the baryon number.  The presence of nonzero $n_L$ biases the sphaleron processes, leading to the production of the baryon asymmetry~\cite{Kuzmin:1985mm}. The expanding bubbles capture the asymmetry, which is preserved assuming the first order EWPT is sufficiently strong so as to quench the sphaleron transitions inside the bubbles.


We compute the CP-asymmetry and resulting $n_L$ using the closed-time path formalism and the ``VEV-insertion" approximation~\cite{Riotto:1998zb,Lee:2004we}. Under these assumptions, one obtains a coupled set of diffusion equations
\begin{eqnarray}
\partial_\mu J^\mu_k = -\sum_{A,j}\, \Gamma_A(\mu_k-\mu_j-\cdots)+S^\mathrm{CPV}_k
\label{eq:diffeq}
\end{eqnarray}
where $J^\mu_k$ is the number density current for species of particle $k$; $\mu_{k,j}$ are chemical potentials for species $k$, $j$ {\em etc}; the $\Gamma_A$ denote rate for particle number changing reactions;  and $S^\mathrm{CPV}_k$ is the CP-violating source term associated with the CP-violating phases and the space-time varying vacuum expectation values of the neutral Higgs scalars. 

In the present instance, the CP-violating source enters the diffusion equations for the new neutral fermions $\chi$ and the neutral component of $\psi$.
Details of the computation for the model under consideration here are similar to those given in see Ref.~\cite{Chao:2014dpa}, so we do not repeat all of them here.
As in obtaining the other curves in Fig. \ref{BAU}, we assume $y_1$ contains the only CP phase and $y_2^\prime =0$. 
The source $S_{\psi}^\mathrm{CPV}=-S_{\chi}^\mathrm{CPV}$ depends crucially on the damping rate of the new fermions, corresponding to the imaginary parts of the poles in the finite temperature fermion propagators~\cite{Kobes:1992ys,Elmfors:1998hh,Enqvist:1997ff}:
\begin{eqnarray}
\label{eq:damping}
\gamma_\psi \sim {3 g^2 + g^{\prime 2 }  + 4 a^2  g_N^2 \over 32 \pi} T \; , \hspace{0.5cm} \gamma_\chi \sim {a^2 g_N^2 \over 8 \pi  } T \; ,
\end{eqnarray} 
where $g_N$ is the gauge coupling of the $U(1)_D$ and $a$ is the relevant hyper-charge of $\psi$ and $\chi$. We have neglected the contribution of the Higgs mediated processes, which is subdominant compared with the contribution from scattering with gauge bosons~\cite{Elmfors:1998hh}.

The expression in Eq.~(\ref{eq:damping}) is general, and one must apply it to the specific representations of interest here. For the $\psi$, one must include the contributions from the SU(2$)_L$, U(1$)_Y$, and U(1$)_D$ gauge bosons when the latter are included in the model to provide for dark matter stability; for the $\chi$ one need only include the new $U(1)_D$ contributions (Higgs mediated interactions contribute to both). Notice that a definitive quantitative treatment of the CPV fermion sources remains an open problem. The VEV insertion approximation used in our calculation is likely to overly estimate the CPV source by at least a factor of  a few. The results given here, thus, provide a conservative basis for evaluating the dark matter and EDM restrictions on the parameter space viable for the EWBG. We refer to \cite{Morrissey:2012db} and references therein for a detailed discussion of the theoretical issues associated with the calculation of the CPV source term.

Solution of the coupled set of diffusion equations (\ref{eq:diffeq}) can be simplified through several observations. First, since all light quarks have nearly equivalent diffusion constants and are mainly produced by strong sphaleron processes, baryon number conservation on time scales shorter than the inverse of the electroweak sphaleron rate approximately implies the relations $q_{1L} =q_{2L}=-2u_R=-2d_R=-2s_R=-2c_R=-2b_R\equiv -2b =  2( Q+T)$. The resulting set of equations  for new fermions  are already given in Ref.~\cite{Chao:2014dpa}. 

Second, one may obtain a set of equations solely for the number densities under the diffusion approximation: ${\vec J} =- D{\vec\bigtriangledown} n$, where $D$ is the diffusion constant for a given species whose number density three-current is  ${\vec J}$.
The diffusion constants for quarks, leptons, and Higgs scalars have been computed previously~\cite{Joyce:1994zn}. Here we compute those for the new fermions, focusing 
elastic scattering, $t$-channel vector boson and Higgs scalar  exchange diagrams that  are expected to dominate the scattering process and play the major role in limiting the diffusion of particles. The diffusion constants mediated by the gauge boson and Higgs bosons can then be written as
\begin{eqnarray}
D_{B^\prime}^{-1} &\approx& {60\over 7 \pi} \alpha_D^2 a^2 T \ln \left(  {32 T^4 \over  M_{B^\prime}^2 }\right)  \; , \\
D_{H_i}^{-1}& \approx & {5 \over 7 \pi } { |y_i ^{} y_i^\prime|^2 +3|y_i^{(\prime) } y_t^{}|^2 \over 16 \pi^2  } T \ln \left(  32 T^2 \over M_{H_i}^2 \right) \;  , 
\end{eqnarray}
The inverse diffusion constants for $\chi$ and $\psi$ are then: $D_\chi^{-1} \approx  D_{B^\prime}^{-1} + \sum_i D_{H_i}^{-1}$  and  $D_\psi^{-1} \approx  {T / 100} + D_{B^\prime}^{-1} + \sum_i D_{H_i}^{-1} $, where  $M_{B^\prime}^2 $ is the one-loop Debye mass of the new gauge boson, $M_{B^\prime}^2 = {4 \pi \over 3 } \alpha_D^{ 2} T^2 $~\cite{Joyce:1994zn,Weldon:1982bn}, where $\alpha_D=g_N^2 /4\pi$. $M_{H_i}^2 $ are the thermal mass of the ``$i$"th Higgs boson. 

We  solve diffusion equations analytically by assuming that strong sphaleron transitions as well as the Yukawa interactions for the top quarks and the new fermions  are all in chemical equilibrium. These assumptions lead to a series of conditions relating the chemical potentials and, therefore, the number densities of various species. Chemical equilibrium for the new fermion Yukawa interactions implies that  ${n_\chi / k_\chi} -{n_H/ k_H} -{n_\psi / k_\psi }=0$, where $n_i$ and $k_i$ $(i=\chi,~\psi,~ H)$ are the number density and the statistical factor for particle ``$i$" respectively.  Furthermore, in the static limit ( where $v_w\to 0$), we have $D_\chi n_\chi =-D_\psi n_\psi$, where we have assumed the boundary conditions $n_\chi(\infty)=n_\chi^\prime(\infty)=n_\psi(\infty)=n_\psi^\prime(\infty)=0$. Therefore, all the charge densities can be expressed as the functions of the Higgs density, and all that remains is to solve for the Higgs density. We refer to Refs.~\cite{Chung:2008aya,Chung:2009cb} for the semi-analytical approximate solution in this case.

The left-handed charge density that biases weak sphaleron transitions, is given by the sum of all charge densities of left-handed quarks and leptons of all generations, $n_L =\sum_i (q_i + \ell_i)$. The left-handed density is converted into a baryon number density $n_B$  through the weak sphaleron process. The following formula describes baryon generation and washout ahead of the bubble wall~\cite{Huet:1995sh}
\begin{eqnarray}
n_B = -n_F {\Gamma_{ws} \over 2 v_w } \int_{-\infty}^0 d z\ n_L (z) e^{ {15\over 4} {\Gamma_{ws} \over v_w} z }
\end{eqnarray}  
where $v_w$ is the bubble wall velocity, $\Gamma_{ws}$ is the weak sphaleron rate and $z$ is the spatial coordinate perpendicular to the wall in the frame where the wall is at rest.  Negative values of $z$ correspond to the symmetric electroweak phase, positive values to the broken phase. 

For explicit numerical illustration, we consider the case wherein stability of the lightest neutral new fermion is guaranteed by the $U(1)_D$ symmetry. In addition, we take the Higgs vev profiles in Ref.\cite{Carena:2000id,Carena:1997gx} and follow Ref. \cite{Lee:2004we,Riotto:1998zb} in our calculation of the CP-conserving relaxation rates for quarks and Higgs.  We assume $k_{Q3} =2 k_T =2 k_B =6$, $k_H =4$, $k_\psi =2k_\chi =2$ and $g_N=0.35$ for numerical calculation.

In Fig.~\ref{BAU}, we show contours of constant $Y_B$ in the $y-\delta$ plane.  Regions between the green dot-dashed lines are consistent with the observed $Y_B$ at $95\%$ C.L..  As noted earlier, the current ACME result places no constraint on the CP-violating phase, so one has adequate parameter space for successful baryogenesis.  However a prospective future search for the electron EDM could probe most of the parameter space.  A similar statement would apply to a future neutron EDM search, assuming an improvement in sensitivity of two orders of magnitude. 
Since $\delta$ dominates the strength of the parity-violating DM-Higgs coupling, the present exclusion limit given by the LUX puts a strong lower bound on the phase. Nevertheless, it is possible to find parameter space where both sizable baryon asymmetry and  a successful dark matter candidate can be derived. We note  that the dark matter/EWBG-viable parameter space varies significantly as one changes inputs of $M_\chi$ and $M_\psi$. A much smaller mass splitting between $M_\chi$ and $M_\psi$  would induce an enhanced source term, {\it via} the resonant enhancement~\cite{Lee:2004we},  for the EWBG,  but introduces a large mixing between two neutral fermions, which may degrade the viability of $\chi$ as a dark matter candidate. A much larger mass splitting would  lead to an inadequate baryon asymmetry on the other hand.

\noindent {\bf Conclusion} The possibility of correlating the abundance of visible and dark matter is a topic of considerable interest, inspiring new paradigms such as asymmetric dark matter~ \cite{Kaplan:2009ag} (for recent reviews, see Refs.~\cite{Petraki:2013wwa,Zurek:2013wia})
or WIMP baryogenesis~\cite{Cui:2011ab}. In this study, we have investigated the possibility that CP-violating, fermionic Higgs portal interactions may both govern the abundance of the dark matter relic density and drive the production of the baryon asymmetry through EWBG while evading the present EDM and dark matter direct detection bounds.  Working with a simple two Higgs doublet model augmented by the presence of vector-like fermions, we illustrated the existence of parameter space regions consistent with the observed abundance of visible and dark matter and present experimental constraints. An extensive study of the dark matter/$Y_B$-viable parameter space, along with the relevant phenomenological implications, will appear in aforthcoming publication.


\noindent{\it  Acknowledgements} 
This work was supported in part by U.S. Department of Energy contract  DE-SC0011095 (MJRM and WCHAO).


\begin{thebibliography}{99}


\bibitem{Ade:2013zuv} 
  P.~A.~R.~Ade {\it et al.}  [Planck Collaboration],
  Astron.\ Astrophys.\  (2014)
  [arXiv:1303.5076 [astro-ph.CO]].
  
\bibitem{Komatsu:2010fb} 
  E.~Komatsu {\it et al.}  [WMAP Collaboration],
  Astrophys.\ J.\ Suppl.\  {\bf 192}, 18 (2011)
  [arXiv:1001.4538 [astro-ph.CO]].
  
  
\bibitem{Patt:2006fw} 
  B.~Patt and F.~Wilczek,
  hep-ph/0605188.
  
\bibitem{Kim:2006af} 
  Y.~G.~Kim and K.~Y.~Lee,
  Phys.\ Rev.\ D {\bf 75}, 115012 (2007)
  [hep-ph/0611069].
  
  
\bibitem{MarchRussell:2008yu} 
  J.~March-Russell, S.~M.~West, D.~Cumberbatch and D.~Hooper,
  JHEP {\bf 0807}, 058 (2008)
  [arXiv:0801.3440 [hep-ph]].

\bibitem{Kim:2008pp} 
  Y.~G.~Kim, K.~Y.~Lee and S.~Shin,
  JHEP {\bf 0805}, 100 (2008)
  [arXiv:0803.2932 [hep-ph]].
  
\bibitem{Ahlers:2008qc} 
  M.~Ahlers, J.~Jaeckel, J.~Redondo and A.~Ringwald,
  Phys.\ Rev.\ D {\bf 78}, 075005 (2008)
  [arXiv:0807.4143 [hep-ph]].
  
  
\bibitem{Feng:2008mu} 
  J.~L.~Feng, H.~Tu and H.~-B.~Yu,
  JCAP {\bf 0810}, 043 (2008)
  [arXiv:0808.2318 [hep-ph]].
  
\bibitem{Andreas:2008xy} 
  S.~Andreas, T.~Hambye and M.~H.~G.~Tytgat,
  JCAP {\bf 0810}, 034 (2008)
  [arXiv:0808.0255 [hep-ph]].
  
\bibitem{Barger:2007im} 
  V.~Barger, P.~Langacker, M.~McCaskey, M.~J.~Ramsey-Musolf and G.~Shaughnessy,
  Phys.\ Rev.\ D {\bf 77}, 035005 (2008)
  [arXiv:0706.4311 [hep-ph]].

  
\bibitem{Barger:2008jx} 
  V.~Barger, P.~Langacker, M.~McCaskey, M.~Ramsey-Musolf and G.~Shaughnessy,
  Phys.\ Rev.\ D {\bf 79}, 015018 (2009)
  [arXiv:0811.0393 [hep-ph]].
  
\bibitem{Kadastik:2009ca} 
  M.~Kadastik, K.~Kannike, A.~Racioppi and M.~Raidal,
  Phys.\ Rev.\ Lett.\  {\bf 104}, 201301 (2010)
  [arXiv:0912.2729 [hep-ph]].
  
\bibitem{Gonderinger:2009jp} 
  M.~Gonderinger, Y.~Li, H.~Patel and M.~J.~Ramsey-Musolf,
  JHEP {\bf 1001}, 053 (2010)
  [arXiv:0910.3167 [hep-ph]].
  
 
\bibitem{Piazza:2010ye} 
  F.~Piazza and M.~Pospelov,
  Phys.\ Rev.\ D {\bf 82}, 043533 (2010)
  [arXiv:1003.2313 [hep-ph]].
  
\bibitem{Arina:2010an} 
  C.~Arina, F.~-X.~Josse-Michaux and N.~Sahu,
  Phys.\ Rev.\ D {\bf 82}, 015005 (2010)
  [arXiv:1004.3953 [hep-ph]].
  
\bibitem{Kanemura:2010sh} 
  S.~Kanemura, S.~Matsumoto, T.~Nabeshima and N.~Okada,
  Phys.\ Rev.\ D {\bf 82}, 055026 (2010)
  [arXiv:1005.5651 [hep-ph]].
  
\bibitem{Englert:2011yb} 
  C.~Englert, T.~Plehn, D.~Zerwas and P.~M.~Zerwas,
  Phys.\ Lett.\ B {\bf 703}, 298 (2011)
  [arXiv:1106.3097 [hep-ph]].
  
\bibitem{Low:2011kp} 
  I.~Low, P.~Schwaller, G.~Shaughnessy and C.~E.~M.~Wagner,
  Phys.\ Rev.\ D {\bf 85}, 015009 (2012)
  [arXiv:1110.4405 [hep-ph]].
  
\bibitem{Djouadi:2011aa} 
  A.~Djouadi, O.~Lebedev, Y.~Mambrini and J.~Quevillon,
  Phys.\ Lett.\ B {\bf 709}, 65 (2012)
  [arXiv:1112.3299 [hep-ph]].
  
  
\bibitem{Kamenik:2012hn} 
  J.~F.~Kamenik and C.~Smith,
  Phys.\ Rev.\ D {\bf 85}, 093017 (2012)
  [arXiv:1201.4814 [hep-ph]].
  
\bibitem{Gonderinger:2012rd} 
  M.~Gonderinger, H.~Lim and M.~J.~Ramsey-Musolf,
  Phys.\ Rev.\ D {\bf 86}, 043511 (2012)
  [arXiv:1202.1316 [hep-ph]].
  
\bibitem{Lebedev:2012zw} 
  O.~Lebedev,
  Eur.\ Phys.\ J.\ C {\bf 72}, 2058 (2012)
  [arXiv:1203.0156 [hep-ph]].
  
\bibitem{LopezHonorez:2012kv} 
  L.~Lopez-Honorez, T.~Schwetz and J.~Zupan,
  Phys.\ Lett.\ B {\bf 716}, 179 (2012)
  [arXiv:1203.2064 [hep-ph]].
  
  
\bibitem{Okada:2012cc} 
  H.~Okada and T.~Toma,
  Phys.\ Lett.\ B {\bf 713}, 264 (2012)
  [arXiv:1203.3116 [hep-ph]].
  
\bibitem{Djouadi:2012zc} 
  A.~Djouadi, A.~Falkowski, Y.~Mambrini and J.~Quevillon,
  Eur.\ Phys.\ J.\ C {\bf 73}, 2455 (2013)
  [arXiv:1205.3169 [hep-ph]].
  
\bibitem{Bai:2012nv} 
  Y.~Bai, V.~Barger, L.~L.~Everett and G.~Shaughnessy,
  Phys.\ Rev.\ D {\bf 88}, no. 1, 015008 (2013)
  [arXiv:1212.5604 [hep-ph]].
  
\bibitem{Englert:2013gz} 
  C.~Englert, J.~Jaeckel, V.~V.~Khoze and M.~Spannowsky,
  JHEP {\bf 1304}, 060 (2013)
  [arXiv:1301.4224 [hep-ph]].
  
\bibitem{Bian:2013wna} 
  L.~Bian, R.~Ding and B.~Zhu,
  Phys.\ Lett.\ B {\bf 728}, 105 (2014)
  [arXiv:1308.3851 [hep-ph]].
  
\bibitem{Chang:2013lfa} 
  C.~-F.~Chang, E.~Ma and T.~-C.~Yuan,
  JHEP {\bf 1403}, 054 (2014)
  [arXiv:1308.6071 [hep-ph], arXiv:1308.6071].
  
\bibitem{Khoze:2013uia} 
  V.~V.~Khoze,
  JHEP {\bf 1311}, 215 (2013)
  [arXiv:1308.6338 [hep-ph]].
  
\bibitem{Okada:2013bna} 
  N.~Okada and O.~Seto,
  Phys.\ Rev.\ D {\bf 89}, 043525 (2014)
  [arXiv:1310.5991 [hep-ph]].
  
\bibitem{Chao:2014ina} 
  W.~Chao,
  arXiv:1412.3823 [hep-ph].
  
\bibitem{Cai:2014hka} 
  Y.~Cai and W.~Chao,
  arXiv:1408.6064 [hep-ph].
  
  
\bibitem{Sakharov:1967dj} 
  A.~D.~Sakharov,
  Pisma Zh.\ Eksp.\ Teor.\ Fiz.\  {\bf 5}, 32 (1967)
  [JETP Lett.\  {\bf 5}, 24 (1967)]
  [Sov.\ Phys.\ Usp.\  {\bf 34}, 392 (1991)]
  [Usp.\ Fiz.\ Nauk {\bf 161}, 61 (1991)].
  
  
\bibitem{Morrissey:2012db} 
  D.~E.~Morrissey and M.~J.~Ramsey-Musolf,
  New J.\ Phys.\  {\bf 14}, 125003 (2012)
  [arXiv:1206.2942 [hep-ph]].
  
\bibitem{Cohen:1993nk} 
  A.~G.~Cohen, D.~B.~Kaplan and A.~E.~Nelson,
  Ann.\ Rev.\ Nucl.\ Part.\ Sci.\  {\bf 43}, 27 (1993)
  [hep-ph/9302210].
  
\bibitem{Trodden:1998ym} 
  M.~Trodden,
  Rev.\ Mod.\ Phys.\  {\bf 71}, 1463 (1999)
  [hep-ph/9803479].


\bibitem{Riotto:1998bt} 
  A.~Riotto,
  hep-ph/9807454.

\bibitem{Riotto:1999yt} 
  A.~Riotto and M.~Trodden,
  Ann.\ Rev.\ Nucl.\ Part.\ Sci.\  {\bf 49}, 35 (1999)
  [hep-ph/9901362].


\bibitem{Quiros:1999jp} 
  M.~Quiros,
  hep-ph/9901312.


\bibitem{Dine:2003ax} 
  M.~Dine and A.~Kusenko,
  Rev.\ Mod.\ Phys.\  {\bf 76}, 1 (2003)
  [hep-ph/0303065].


\bibitem{Cline:2006ts} 
  J.~M.~Cline,
  hep-ph/0609145.
  
\bibitem{Engel:2013lsa} 
  J.~Engel, M.~J.~Ramsey-Musolf and U.~van Kolck,
  Prog.\ Part.\ Nucl.\ Phys.\  {\bf 71}, 21 (2013)
  [arXiv:1303.2371 [nucl-th]].
  
\bibitem{Pospelov:2005pr} 
  M.~Pospelov and A.~Ritz,
  Annals Phys.\  {\bf 318}, 119 (2005)
  [hep-ph/0504231].


\bibitem{Aprile:2012nq} 
  E.~Aprile {\it et al.}  [XENON100 Collaboration],
  Phys.\ Rev.\ Lett.\  {\bf 109}, 181301 (2012)
  [arXiv:1207.5988 [astro-ph.CO]].
  
\bibitem{Akerib:2013tjd} 
  D.~S.~Akerib {\it et al.}  [LUX Collaboration],
  arXiv:1310.8214 [astro-ph.CO].


\bibitem{Branco:2011iw} 
  G.~C.~Branco, P.~M.~Ferreira, L.~Lavoura, M.~N.~Rebelo, M.~Sher and J.~P.~Silva,
  Phys.\ Rept.\  {\bf 516}, 1 (2012)
  [arXiv:1106.0034 [hep-ph]].
  
  
\bibitem{Chao:2014dpa} 
  W.~Chao and M.~J.~Ramsey-Musolf,
  arXiv:1406.0517 [hep-ph].
  
  \bibitem{2hdmpdm}
  W. ~Chao, H. ~Guo and M.~J.~Ramsey-Musolf, to appear.
  
\bibitem{Bertone:2004pz} 
  G.~Bertone, D.~Hooper and J.~Silk,
  Phys.\ Rept.\  {\bf 405}, 279 (2005)
  [hep-ph/0404175].
  
  
   
 \bibitem{Barr:1990vd} 
  S.~M.~Barr and A.~Zee,
  Phys.\ Rev.\ Lett.\  {\bf 65}, 21 (1990)
  [Erratum-ibid.\  {\bf 65}, 2920 (1990)].
  
 
\bibitem{Giudice:2005rz} 
  G.~F.~Giudice and A.~Romanino,
  Phys.\ Lett.\ B {\bf 634}, 307 (2006)
  [hep-ph/0510197].
  
  
\bibitem{Baron:2013eja} 
  J.~Baron {\it et al.}  [ACME Collaboration],
  Science {\bf 343}, no. 6168, 269 (2014)
  [arXiv:1310.7534 [physics.atom-ph]].
  
  
\bibitem{Baker:2006ts} 
  C.~A.~Baker, D.~D.~Doyle, P.~Geltenbort, K.~Green, M.~G.~D.~van der Grinten, P.~G.~Harris, P.~Iaydjiev and S.~N.~Ivanov {\it et al.},
  Phys.\ Rev.\ Lett.\  {\bf 97}, 131801 (2006)
  [hep-ex/0602020].
  
  
  \bibitem{Li:2008kz} 
  Y.~Li, S.~Profumo and M.~Ramsey-Musolf,
  Phys.\ Rev.\ D {\bf 78}, 075009 (2008)
  [arXiv:0806.2693 [hep-ph]].
  
 \bibitem{Li:2008ez} 
  Y.~Li, S.~Profumo and M.~Ramsey-Musolf,
  Phys.\ Lett.\ B {\bf 673}, 95 (2009)
  [arXiv:0811.1987 [hep-ph]].

  
    
\bibitem{Cline:1996mga} 
  J.~M.~Cline and P.~-A.~Lemieux,
  Phys.\ Rev.\ D {\bf 55}, 3873 (1997)
  [hep-ph/9609240].
  
  
\bibitem{Fromme:2006cm} 
  L.~Fromme, S.~J.~Huber and M.~Seniuch,
  JHEP {\bf 0611}, 038 (2006)
  [hep-ph/0605242].
  
\bibitem{Dorsch:2013wja} 
  G.~C.~Dorsch, S.~J.~Huber and J.~M.~No,
  JHEP {\bf 1310}, 029 (2013)
  [arXiv:1305.6610 [hep-ph]].
  
  
   
\bibitem{Kuzmin:1985mm} 
  V.~A.~Kuzmin, V.~A.~Rubakov and M.~E.~Shaposhnikov,
  Phys.\ Lett.\ B {\bf 155}, 36 (1985).
  
  
  
\bibitem{Lee:2004we} 
  C.~Lee, V.~Cirigliano and M.~J.~Ramsey-Musolf,
  Phys.\ Rev.\ D {\bf 71}, 075010 (2005)
  [hep-ph/0412354].
  
   
\bibitem{Riotto:1998zb}
  A.~Riotto,
  Phys.\ Rev.\ D {\bf 58} (1998) 095009
  [hep-ph/9803357].
  
  
\bibitem{Kobes:1992ys} 
  R.~Kobes, G.~Kunstatter and K.~Mak,
  Phys.\ Rev.\ D {\bf 45}, 4632 (1992).
  
\bibitem{Elmfors:1998hh} 
  P.~Elmfors, K.~Enqvist, A.~Riotto and I.~Vilja,
  Phys.\ Lett.\ B {\bf 452}, 279 (1999)
  [hep-ph/9809529].
  
\bibitem{Enqvist:1997ff} 
  K.~Enqvist, A.~Riotto and I.~Vilja,
  Phys.\ Lett.\ B {\bf 438}, 273 (1998)
  [hep-ph/9710373].
  
  
\bibitem{Joyce:1994zn} 
  M.~Joyce, T.~Prokopec and N.~Turok,
  Phys.\ Rev.\ D {\bf 53}, 2930 (1996)
  [hep-ph/9410281].
  
\bibitem{Weldon:1982bn} 
  H.~A.~Weldon,
  Phys.\ Rev.\ D {\bf 26}, 2789 (1982).
  
  

\bibitem{Chung:2008aya} 
  D.~J.~H.~Chung, B.~Garbrecht, M.~J.~Ramsey-Musolf and S.~Tulin,
  Phys.\ Rev.\ Lett.\  {\bf 102}, 061301 (2009)
  [arXiv:0808.1144 [hep-ph]].
  
\bibitem{Chung:2009cb} 
  D.~J.~H.~Chung, B.~Garbrecht, M.~J.~Ramsey-Musolf and S.~Tulin,
  Phys.\ Rev.\ D {\bf 81}, 063506 (2010)
  [arXiv:0905.4509 [hep-ph]].
  
  
\bibitem{Huet:1995sh} 
  P.~Huet and A.~E.~Nelson,
  Phys.\ Rev.\ D {\bf 53}, 4578 (1996)
  [hep-ph/9506477].

\bibitem{Carena:2000id} 
  M.~S.~Carena, J.~M.~Moreno, M.~Quiros, M.~Seco and C.~E.~M.~Wagner,
  Nucl.\ Phys.\ B {\bf 599}, 158 (2001)
  [hep-ph/0011055].
  
  
\bibitem{Carena:1997gx} 
  M.~S.~Carena, M.~Quiros, A.~Riotto, I.~Vilja and C.~E.~M.~Wagner,
  Nucl.\ Phys.\ B {\bf 503}, 387 (1997)
  [hep-ph/9702409].
  
  
  

  
 
\bibitem{Kaplan:2009ag} 
  D.~E.~Kaplan, M.~A.~Luty and K.~M.~Zurek,
  Phys.\ Rev.\ D {\bf 79}, 115016 (2009)
  [arXiv:0901.4117 [hep-ph]].
    
\bibitem{Petraki:2013wwa} 
  K.~Petraki and R.~R.~Volkas,
  Int.\ J.\ Mod.\ Phys.\ A {\bf 28}, 1330028 (2013)
  [arXiv:1305.4939 [hep-ph]].
  
\bibitem{Zurek:2013wia} 
  K.~M.~Zurek,
  Phys.\ Rept.\  {\bf 537}, 91 (2014)
  [arXiv:1308.0338 [hep-ph]].

  
\bibitem{Cui:2011ab} 
  Y.~Cui, L.~Randall and B.~Shuve,
  JHEP {\bf 1204}, 075 (2012)
  [arXiv:1112.2704 [hep-ph]].
  

\end{thebibliography}
\end{document}